\documentclass[oneside,english]{amsart}
\usepackage[T1]{fontenc}
\usepackage[latin9]{inputenc}
\usepackage{textcomp}
\usepackage{amstext}
\usepackage{amsthm}
\usepackage{amssymb}

\makeatletter

\providecommand{\tabularnewline}{\\}

\numberwithin{equation}{section}
\numberwithin{figure}{section}
\theoremstyle{plain}
\newtheorem{thm}{\protect\theoremname}
  \theoremstyle{plain}
  \newtheorem{lem}[thm]{\protect\lemmaname}

\usepackage{babel}

\usepackage{babel}
\providecommand{\lemmaname}{Lemma}
\providecommand{\theoremname}{Theorem}

\makeatother

\usepackage{babel}
  \providecommand{\lemmaname}{Lemma}
\providecommand{\theoremname}{Theorem}

\begin{document}

\title{Right-handed neutrinos and $U\left(1\right)_{X}$ symmetry-breaking}

\author{C. Herbert Clemens and Stuart Raby}

\address{Mathematics Dept./Physics Dept., Ohio State University, Columbus
OH 43210}

\date{February 27, 2020}
\begin{abstract}
The authors have proposed a global model for Heterotic $F$-theory
duality with Wilson line symmetry-breaking and a $4+1$ split of the
$F$-theory spectral divisor. Goals of this note are to treat the
existence of right-handed neutrinos in our $F$-theory model, show
that the $\mathbb{Z}_{2}$-action in our model breaks the $U\left(1\right)_{X}$-symmetry
associated to the $4+1$ split to $\mathbb{Z}_{2}$-matter parity,
and to identify Yukawa couplings for the MSSM matter fields.
\end{abstract}

\email{clemens.43@osu.edu, raby.1@osu.edu}

\maketitle
In \cite{Clemens-3} we proposed a model for Heterotic $F$-theory
duality with Wilson line symmetry-breaking and a $4+1$ split of the
$F$-theory spectral divisor. One goal of this note is to call attention
to the existence of right-handed neutrinos in our $F$-theory model.
As pointed out in §4 of \cite{Blumenhagen} such existence may be
evidence for the $U\left(1\right)_{X}$-symmetry that remains after
the Higgsing of $E_{8}$ via 
\[
E_{8}\Rightarrow SU\left(5\right)_{gauge}\oplus\left[SU\left(4\right)\oplus U\left(1\right)_{X}\right]_{Higgs}
\]
occasioned by the $4+1$ split of the spectral divisor. In addition,
as a result of the $\mathbb{Z}_{2}$-action that supports the Wilson
line we argue that the $U\left(1\right)_{X}$-symmetry is, in fact,
broken to $\mathbb{Z}_{2}$-matter parity. Finally we identify co-dimension
$3$ singularities which determine Yukawa couplings for the MSSM matter
fields.

\section{The geometric model}

The Tate form for the Calabi-Yau fourfold $W_{4}/B_{3}$ in our set-up
is given by the equation 
\begin{equation}
\left|\begin{array}{cc}
x^{3}+a_{4}zwx^{2}+a_{2}z^{3}w^{2}x+a_{0}z^{5}w^{3} & 1\\
wy^{2}-\left(a_{5}wx+a_{3}z^{2}w^{2}\right)y & 1
\end{array}\right|=0\label{eq:T}
\end{equation}
over a Fano threefold 
\[
B_{3}=\mathbb{P}_{\left[u_{0},v_{0}\right]}\times D_{2}
\]
where $D_{2}$ is a special del Pezzo surface with a $\mathbb{Z}_{4}$-symmetry.
The pair $W_{4}/B_{3}$ admits an equivariant $\mathbb{Z}_{2}$-action
with respect to which 
\[
a_{j},\,z,\,\frac{y}{x}=t\in H^{0}\left(K_{B_{3}}^{-1}\right)^{\left[-1\right]},
\]
the skew eigenspace.

\subsection{The second section}

We impose the condition 
\begin{equation}
a_{02345}:=a_{0}+a_{2}+a_{3}+a_{4}+a_{5}=0\label{eq:secsec}
\end{equation}
from which one checks that (\ref{eq:T}) is satisfied on the locus
\[
\begin{array}{c}
x=z^{2}w\\
y=z^{3}w
\end{array}
\]
so that, besides the tautological section 
\[
\zeta:=\left\{ \left[w,x,y\right]=\left[0,0,1\right]\right\} 
\]
of $\tilde{W}_{4}/B_{3}$ we have a second section 
\[
\tau:=\left\{ \left[w,x,y\right]=\left[w,z^{2}w,z^{3}w\right]\right\} .
\]
These two sections generate a third section obtained for each $b_{3}\in B_{3}$
from the third point of intersection 
\[
\upsilon:=\left\{ \left[w,x,y\right]=\left[w,z^{2}w,-\left(z+a_{420}\right)z^{2}w\right]\right\} 
\]
of the line between $\zeta\left(b_{3}\right)$ and $\tau\left(b_{3}\right)$
with the elliptic fiber of $\tilde{W}_{4}/B_{3}$ over $b_{3}$. (Again
$a_{420}:=a_{4}+a_{2}+a_{0}$.)

\subsection{The singular locus}

By Bertini's theorem, the singular locus of the Tate form lies in
\[
y^{2}-x^{3}=\left|\begin{array}{cc}
a_{4}zwx^{2}+a_{2}z^{3}w^{2}x+a_{0}z^{5}w^{3} & 1\\
-\left(a_{5}wx+a_{3}z^{2}w^{2}\right)y & 1
\end{array}\right|=0.
\]
Parametrizing $\left\{ y^{2}-x^{3}=0\right\} $ by 
\[
x=t^{2}w,\,y=t^{3}w
\]
the pull-back of the Tate form to $\left\{ y^{2}-x^{3}=0\right\} $
is given by 
\[
\begin{array}{c}
a_{5}t^{5}+a_{4}t^{4}z+a_{3}t^{3}z^{2}+a_{2}t^{2}z^{3}+a_{0}z^{5}=\\
\left(a_{5}t^{4}+a_{54}t^{3}z-a_{20}t^{2}z^{2}-a_{0}z^{3}\left(t+z\right)\right)\left(t-z\right)=0.
\end{array}
\]

The subvariety 
\[
\mathcal{D}^{\left(4\right)}+\mathcal{D}^{\left(1\right)}\subseteq\left\{ y^{2}-x^{3}=0\right\} 
\]
is called the spectral variety. We make the assumption that 
\[
\left\{ u_{0}v_{0}=z=0\right\} \subseteq\left\{ a_{5}=z=0\right\} 
\]
so that 
\[
\left\{ u_{0}v_{0}=z=0\right\} \subseteq\mathcal{D}^{\left(4\right)}.
\]

\subsection{Double-cover model}

As in §6.1 of \cite{Clemens-3}, we project from the third section
$\upsilon$ mentioned above and write the birational model $\bar{W}_{4}/B_{3}$
of $W_{4}/B_{3}$ as a branched double cover of 
\[
Q:=\mathbb{P}\left(\mathcal{O}_{B_{3}}\oplus\mathcal{O}_{B_{3}}\left(N\right)\right)
\]
where $Q/B_{3}$ has affine fiber coordinate $\vartheta_{0}$. The
branch locus $\Delta$ is then given by the equation

\begin{equation}
\begin{array}{c}
0=\left(\left(3z^{2}+a_{4}z\right)-\vartheta_{0}\left(\vartheta_{0}-a_{5}\right)\right)^{2}\\
-4\left(\left(3z^{4}+\left(2a_{4}+a_{2}-a_{5}\right)z^{3}-a_{5}a_{420}z^{2}\right)+\left(2z^{3}+a_{420}z^{2}\right)\vartheta_{0}\right)
\end{array}\label{eq:affdelta}
\end{equation}
on the space 
\[
Q-\left\{ X=0\right\} =\left|\mathcal{O}_{B_{3}}\left(N\right)\right|,
\]
the total space of the line bundle $\mathcal{O}_{B_{3}}\left(N\right)$.

\subsection{Image of spectral divisor}

As in §7.4 of \cite{Clemens-3}, to compute the image 
\[
\mathcal{C}_{0}^{\left(4\right)}+\mathcal{C}_{0}^{\left(1\right)}\subseteq Q
\]
of $\mathcal{D}^{\left(4\right)}+\mathcal{D}^{\left(1\right)}$ we
write 
\[
\begin{array}{c}
t=\frac{\vartheta_{0}-\left(z+a_{420}\right)}{2}\\
t-z=\frac{\vartheta_{0}-\left(3z+a_{420}\right)}{2}\\
t+z=\frac{\vartheta_{0}+z-a_{420}}{2}
\end{array}
\]
so that the equations of the images of the two components are 
\begin{equation}
\begin{array}{c}
\left(\vartheta_{0}-\left(z+a_{420}\right)\right)^{2}\left(\left(a_{5}\left(\vartheta_{0}-\left(z+a_{420}\right)\right)+2a_{4}z\right)\left(\vartheta_{0}+z-a_{420}\right)-4a_{420}z^{2}\right)\\
-8a_{0}z^{3}\left(\vartheta_{0}+z-a_{420}\right)=0
\end{array}\label{eq:speceq}
\end{equation}
and 
\[
\vartheta_{0}-\left(3z+a_{420}\right)=0.
\]

In \cite{Clemens-3} we show that $\mathcal{C}_{0}^{\left(4\right)}+\mathcal{C}_{0}^{\left(1\right)}$
lifts isomorphically to the image of $\mathcal{D}^{\left(4\right)}+\mathcal{D}^{\left(1\right)}$
in the branched double cover $\bar{W}_{4}/Q$. \eqref{eq:speceq}
implies that the map from the spectral divisor to $Q$ is birational
so that the inverse image of both $\mathcal{C}_{0}^{\left(4\right)}$
and $\mathcal{C}_{0}^{\left(1\right)}$ in $\overline{W}_{4}$ are
reducible and only one of their two components correspond to the image
\begin{equation}
\mathcal{\bar{C}}_{Higgs}=\mathcal{\bar{C}}_{Higgs}^{\left(4\right)}+\mathcal{\bar{C}}_{Higgs}^{\left(1\right)}\subseteq\overline{W}_{4}\label{eq:specHiggs}
\end{equation}
of $\mathcal{D}^{\left(4\right)}+\mathcal{D}^{\left(1\right)}.$ Here
the divisors $\left(\zeta\right)$ and $\left(\tau\right)$ given
by the two sections each project to the locus $\left\{ \vartheta_{0}=\infty\right\} \subseteq Q$
and $\mathcal{C}_{0}^{\left(4\right)}$ the lift $\mathcal{\bar{C}}_{Higgs}^{\left(4\right)}$
is the component intersecting the proper transform of the section
$\left(\tau\right)$ .

Furthermore, the first two modifications $W_{4}^{\left(1\right)}$
and $W_{4}^{\left(2\right)}$ of $\bar{W}_{4}/Q$ are induced via
fibered product from modifications $Q^{\left(1\right)}$ and $Q^{\left(2\right)}$
of $Q$ so it will be convenient to describe these modifications in
terms of their effect on the branch locus $\Delta\subseteq Q$, modifications
that we denote as $\Delta^{\left(1\right)}$ and $\Delta^{\left(2\right)}=\tilde{\Delta}$.

Also, over the divisor $S_{\mathrm{GUT}}:=\left\{ z=0\right\} \subseteq B_{3}$,
the spectral locus $\left(\mathcal{C}_{0}^{\left(4\right)}+\mathcal{C}_{0}^{\left(1\right)}\right)\times_{B_{3}}S_{\mathrm{GUT}}$
has afffine equations 
\begin{equation}
\begin{array}{c}
a_{5}\left(\vartheta_{0}-a_{420}\right)^{4}=0\\
\vartheta_{0}-a_{420}=0
\end{array}\label{eq:sGUT}
\end{equation}
in $Q\times_{B_{3}}S_{\mathrm{GUT}}$ whereas the affine equation
for $\Delta\times_{B_{3}}S_{\mathrm{GUT}}$ in $Q\times_{B_{3}}S_{\mathrm{GUT}}$
is 
\[
\vartheta_{0}^{2}\left(\vartheta_{0}-a_{5}\right)^{2}=0.
\]
Finally we lift $S_{\mathrm{GUT}}$ to 
\[
\bar{S}_{\mathrm{GUT}}:=\left\{ z=\vartheta_{0}-a_{420}=0\right\} \subseteq\mathcal{\bar{C}}_{Higgs}^{\left(4\right)}\subseteq\bar{W}_{4}.
\]

\subsection{Resolution of singularities of $\bar{W}_{4}$}

\subsubsection{Resolution of codimension-one singularities of $\Delta$}

Partial resolution of singularities of $\Delta$ is achieved by blowing
up smooth loci in $Q$. The proper transform $\Delta^{\left(1\right)}$
of $\Delta$ in the first modification $Q^{\left(1\right)}$ of $Q$
is given by the equation 
\begin{equation}
\begin{array}{c}
0=\left(a_{5}\vartheta_{1}+a_{4}Z_{14}\right)^{2}+4a_{5}a_{420}Z_{14}^{2}\\
+\left(-2a_{5}\vartheta_{1}^{3}-2a_{4}\vartheta_{1}^{2}Z_{14}+\left(6a_{5}-4a_{420}\right)\vartheta_{1}Z_{14}^{2}-\left(2a_{4}+4a_{2}-4a_{5}\right)Z_{14}^{3}\right)Z_{0}\\
+\left(\vartheta_{1}^{4}-6\vartheta_{1}^{2}Z_{14}^{2}-8Z_{14}^{3}\vartheta_{1}-3Z_{14}^{4}\right)Z_{0}^{2}.
\end{array}\label{eq:DSp-1}
\end{equation}

A second modification yields $\Delta^{\left(2\right)}\subseteq Q^{\left(2\right)}=\tilde{Q}$
given by the equation

\begin{equation}
\begin{array}{c}
0=\left(\vartheta_{1}\vartheta_{2}+a_{4}\tilde{Z}_{14}\right)^{2}\\
+\left(\left(6a_{5}+4a_{420}\right)\vartheta_{2}-\left(4a_{5}+2a_{4}+4a_{2}\right)Z_{0}\tilde{Z}_{14}\right)\tilde{Z}_{14}^{2}Z_{23}\\
-\left(6\vartheta_{2}^{2}-8\vartheta_{2}Z_{0}\tilde{Z}_{14}+3\tilde{Z}_{14}^{2}Z_{0}^{2}\right)\tilde{Z}_{14}^{2}Z_{23}^{2}.
\end{array}\label{eq:delta2}
\end{equation}
However $\Delta^{\left(2\right)}$ still has a singular locus, namely
a smooth one-dimensional locus that becomes a smooth curve $C_{\mathbf{\bar{5}}}^{\left(44\right)}$
of nodal singularities in the double cover $W_{4}^{\left(2\right)}$.

\subsubsection{Resolution of codimension-two singularities of $\Delta^{\left(2\right)}$}

It becomes important at this point that 
\[
S_{\mathrm{GUT}}:=\left\{ z=0\right\} \subseteq B_{3}
\]
is a smooth $K3$-surface on which the $\mathbb{Z}_{2}$-action is
free, that the imbedded image $\Sigma_{\mathbf{\bar{5}}}^{\left(44\right)}\subseteq B_{3}$
of the curve $C_{\mathbf{\bar{5}}}^{\left(44\right)}\subseteq\tilde{Q}$
lies in $S_{\mathrm{GUT}}$, and that $C_{\mathbf{\bar{5}}}^{\left(44\right)}$
is disjoint from (the proper transform of) $\mathcal{C}_{0}^{\left(4\right)}+\mathcal{C}_{0}^{\left(1\right)}$.
A choice of either of the two possible small, therefore crepant, resolutions
of the nodal locus 
\[
C_{\mathbf{\bar{5}}}^{\left(44\right)}\subseteq W_{4}^{\left(2\right)}
\]
yields the crepant resolution $\tilde{W}_{4}/B_{3}$ that is our $F$-theory
model.

\subsubsection{Lifting $S_{\mathrm{GUT}}\subseteq B_{3}$}

We choose the intersection of the proper transform of $\left\{ \vartheta_{0}-a_{420}=0\right\} $
in $Q^{\left(1\right)}$ to support the lifting $S_{\mathrm{GUT}}^{\left(1\right)}$
of $S_{\mathrm{GUT}}\subseteq B_{3}$ into $Q^{\left(1\right)}$ .
In the notation of \cite{Clemens-3} the equation of $S_{\mathrm{GUT}}^{\left(1\right)}$
can be variously written as 
\[
Z_{14}=\vartheta_{1}Z_{0}-a_{420}=0
\]
A careful study of the first modification in \cite{Clemens-3} shows
that the first lifting $S_{\mathrm{GUT}}^{\left(1\right)}$ of $S_{\mathrm{GUT}}$
into $S_{\mathrm{GUT}}\times_{B_{3}}W_{4}^{\left(1\right)}$ is an
isomorphism with residual component $\left\{ Z_{0}=a_{420}=0\right\} $.
Since $S_{\mathrm{GUT}}^{\left(1\right)}$ maps into $\left\{ Z_{14}=0\right\} $
and meets $\left\{ Z_{0}=0\right\} $ only over $\left\{ a_{420}=z=0\right\} $.
That is, it must lie in $D_{4}$ where it has equation 
\[
\vartheta_{1}Z_{0}-a_{420}=0
\]
and contains liftings of matter curves. Similarly, the proper lifting
$\left\{ \left\{ a_{5}-\vartheta_{1}Z_{0}\right\} =Z_{14}=0\right\} $
of $\left\{ a_{5}-\vartheta_{0}\right\} $ has residual component
$\left\{ Z_{0}=a_{5}=0\right\} $ meeting $\left\{ Z_{0}=0\right\} $
only over $\left\{ a_{5}=z=0\right\} $. $S_{\mathrm{GUT}}^{\left(1\right)}$
meets the proper transform $\left\{ a_{5}-\vartheta_{1}Z_{0}=Z_{14}=0\right\} $
where 
\[
a_{5}-\vartheta_{1}Z_{0}=\vartheta_{1}Z_{0}-a_{420}=Z_{14}=0,
\]
that is, along the entire locus 
\[
\left\{ a_{420}=a_{5}=Z_{14}=0\right\} .
\]
So we obtain the isomorphic lifting $S_{\mathrm{GUT}}^{\left(2\right)}\subseteq W_{4}^{\left(2\right)}$.
Finally, since $S_{\mathrm{GUT}}^{\left(2\right)}$ maps into $\left\{ \tilde{Z}_{14}=0\right\} $
and so never meets the singular curve $C_{\mathbf{\bar{5}}}^{\left(44\right)}$,
it lifts isomorphically to 
\[
\tilde{S}_{\mathrm{GUT}}\subseteq D_{4}\subseteq\tilde{W}_{4}.
\]
Furthermore 
\begin{equation}
a_{5}=\vartheta_{1}Z_{0}+\vartheta_{2}Z_{23}.\label{eq:6}
\end{equation}

\subsection{Matter and Higgs curves in the spectral divisor $\mathcal{\tilde{C}}_{Higgs}^{\left(4\right)}\subseteq\tilde{W}_{4}$}

We have let 
\[
\mathcal{C}_{Higgs}=\mathcal{C}_{Higgs}^{\left(4\right)}+\mathcal{C}_{Higgs}^{\left(1\right)}\subseteq\tilde{Q}=Q^{\left(2\right)}
\]
denote the proper (also the total) transform of \eqref{eq:specHiggs}
and 
\[
\mathcal{\tilde{C}}_{Higgs}\subseteq\tilde{W}_{4}
\]
denote the proper (also the total) transform of $\mathcal{\bar{C}}_{Higgs}$.

The matter curves are given by 
\begin{equation}
\Sigma_{\mathbf{10}}^{\left(4\right)}:=\left\{ a_{5}=\tilde{Z}_{14}=\vartheta_{1}\tilde{Z}_{0}Z_{23}-a_{420}=0\right\} \subseteq\mathcal{\tilde{C}}_{Higgs}^{\left(4\right)}\label{eq:intinfo}
\end{equation}
where $a_{5}=0$ implies that $Z_{0}=\tilde{Z}_{0}Z_{23}$ (see Section
\ref{sec:Configuration-of-components}) and

\begin{equation}
\Sigma_{\mathbf{\bar{5}}}^{\left(41\right)}:=\left\{ Z_{0}=a_{420}=\tilde{Z}_{14}=0\right\} \subseteq\mathcal{\tilde{C}}_{Higgs}^{\left(4\right)}.\label{eq:intinfo'}
\end{equation}
The Higgs curve is given by 
\begin{equation}
\Sigma_{\mathbf{\bar{5}}}^{\left(44\right)}:=\left\{ \left|\begin{array}{cc}
a_{4} & a_{5}\\
a_{0}+a_{3} & -a_{3}
\end{array}\right|=\tilde{Z}_{14}=\vartheta_{1}Z_{0}-a_{420}=0\right\} \subseteq\mathcal{\tilde{C}}_{Higgs}^{\left(4\right)}.\label{eq:sigma44}
\end{equation}
All three curves lie in the surface $\tilde{S}_{\mathrm{GUT}}$ given
in $D_{4}$. The identity 
\[
a_{5}=\vartheta_{1}Z_{0}+\vartheta_{2}Z_{23}
\]
can be used to give alternative formulas for the matter and Higgs
curves by substituting $a_{5}-\vartheta_{2}Z_{23}$ for $\vartheta_{1}Z_{0}$
in \eqref{eq:sigma44} or for $\vartheta_{1}\tilde{Z}_{0}Z_{23}$
in \eqref{eq:intinfo}.

\section{Configuration of components of \label{sec:Configuration-of-components}$\tilde{W}_{4}\times_{B_{3}}S_{\mathrm{GUT}}$ }

Since $W_{4}^{\left(2\right)}/Q^{\left(2\right)}$ is a branched double
cover we will first describe the three components of $Q^{\left(2\right)}\times_{B_{3}}S_{\mathrm{GUT}}$.
These are $\left\{ \tilde{Z}_{14}=0\right\} $ that is the proper
transform of $Q\times_{B_{3}}S_{\mathrm{GUT}}$, $\left\{ Z_{0}=0\right\} $that
is the proper transform of the exceptional locus of the blow-up of
the locus $\left\{ \vartheta_{0}=z=0\right\} $ in $Q$, and $\left\{ Z_{23}=0\right\} $,
the exceptional locus of the blow-up of the locus $\left\{ a_{5}-\vartheta_{1}Z_{0}=Z_{14}=0\right\} $
in $Q^{\left(1\right)}$. Over a general $p\in S_{\mathrm{GUT}}$,
the configuration is a chain 
\begin{equation}
\left\{ Z_{0}\left(p\right)=0\right\} \cup\left\{ \tilde{Z}_{14}\left(p\right)=0\right\} \cup\left\{ Z_{23}\left(p\right)=0\right\} \label{eq:c1}
\end{equation}
where the branch locus \eqref{eq:delta2} does not intersect $\left\{ \tilde{Z}_{14}=0\right\} $
since $\vartheta_{1}\vartheta_{2}$ and $\tilde{Z}_{14}$ have no
common zeros, and intersects $\left\{ Z_{23}=0\right\} $ when $\left(\vartheta_{1}\frac{\vartheta_{2}}{\tilde{Z}_{14}}+a_{4}\right)^{2}=0$.

If $a_{5}\text{\ensuremath{\neq}}0$ the centers of the first and
second modifications are disjoint, so 
\[
\left\{ Z_{0}=0\right\} \cap\left\{ Z_{23}=0\right\} \cap\left\{ a_{5}\text{\ensuremath{\neq}}0\right\} =\textrm{Ø}
\]
and $\left\{ \vartheta_{1}=0\right\} $ meets only $\left\{ Z_{0}=0\right\} $
where it meets transversely as a smooth section disjoint from $\left\{ \tilde{Z}_{14}=0\right\} \cup\left\{ Z_{23}=0\right\} $.
After only the first modification, the equation of the branch locus
restricted to $\left\{ Z_{0}=0\right\} $ is $\left(a_{5}\frac{\vartheta_{1}}{Z_{14}}+a_{4}\right)^{2}+4a_{5}a_{420}=0$,
so, again as long as $a_{5}\text{\ensuremath{\neq}}0$, $\left\{ Z_{0}=0\right\} $
has a double cover $D_{0}$ simply branched at two distinct points
that come together when additionally $a_{420}=0$. Thus if $a_{420}=0$
\[
D_{0}=D_{01}+D_{40}
\]
indicating attachment of components to $D_{1}$ and $D_{4}$ respectively.

If $a_{5}\left(p\right)=0$, the proper transform of $\left\{ Z_{0}\left(p\right)\subseteq Q^{\left(1\right)}\left(p\right)\right\} $
does not equal its total transform that becomes $\left\{ \tilde{Z}_{0}\left(p\right)Z_{23}\left(p\right)=0\right\} \subseteq Q^{\left(2\right)}\left(p\right)$.
Thus from \eqref{eq:6} we have 
\[
\vartheta_{1}\left(p\right)\tilde{Z}_{0}\left(p\right)Z_{23}\left(p\right)+\vartheta_{2}\left(p\right)Z_{23}\left(p\right)=0
\]
giving the functional equation 
\[
\vartheta_{2}=-\vartheta_{1}\tilde{Z}_{0}
\]
on $Q^{\left(2\right)}\times_{B_{3}}\left\{ a_{5}=0\right\} $. $\tilde{Z}_{0}$
divides $\vartheta_{2}$ when $a_{5}=0$ therefore 
\begin{equation}
\left\{ \tilde{Z}_{14}\left(p\right)=0\right\} \cap\left\{ \tilde{Z}_{0}\left(p\right)=0\right\} =\textrm{Ø}.\label{eq:14}
\end{equation}
So the chain configuration \eqref{eq:c1} changes, namely it becomes
\[
\left\{ \tilde{Z}_{14}\left(p\right)=0\right\} \cup\left\{ Z_{23}\left(p\right)=0\right\} \cup\left\{ \tilde{Z}_{0}\left(p\right)=0\right\} 
\]
Also when $a_{5}=0,$ \eqref{eq:delta2} becomes 
\begin{equation}
\begin{array}{c}
0=\left(-\vartheta_{1}^{2}\tilde{Z}_{0}+a_{4}\tilde{Z}_{14}\right)^{2}\\
-\left(4a_{420}\vartheta_{1}+\left(2a_{4}+4a_{2}\right)Z_{23}\tilde{Z}_{14}\right)\tilde{Z}_{0}\tilde{Z}_{14}^{2}Z_{23}\\
-\left(6\vartheta_{1}^{2}+8\vartheta_{1}\tilde{Z}_{14}+3\tilde{Z}_{14}^{2}\right)\tilde{Z}_{0}^{2}\tilde{Z}_{14}^{2}Z_{23}^{2}.
\end{array}\label{eq:15}
\end{equation}
Therefore by \eqref{eq:14} the branch locus $\Delta^{\left(2\right)}=\tilde{\Delta}$
does not intersect $\left\{ \tilde{Z}_{14}\left(p\right)=0\right\} $
and it intersects $\left\{ \tilde{Z}_{0}\left(p\right)=0\right\} $
only if additionally $a_{4}=0$, in which case its intersection is
the entire fiber $\left\{ \tilde{Z}_{0}^{2}\left(p\right)=0\right\} $.
$\Delta^{\left(2\right)}=\tilde{\Delta}$ intersects $\left\{ Z_{23}\left(p\right)=0\right\} $
when $\left(-\vartheta_{1}^{2}\tilde{Z}_{0}+a_{4}\tilde{Z}_{14}\right)^{2}=0$.
So if in addition $a_{4}\left(p\right)=0$ this intersection is given
by $Z_{23}\left(p\right)=\tilde{Z}_{0}^{2}\left(p\right)=0$. From
§7.2.3 of \cite{Clemens-3} the equation for $W_{4}^{\left(2\right)}\cap\left\{ a_{5}=0\right\} $
becomes 
\[
\left|\begin{array}{cc}
w_{2}-\left(a_{4}\tilde{Z}_{14}-\vartheta_{1}^{2}\tilde{Z}_{0}\right) & \tilde{Z}_{14}^{2}Z_{23}\\
\tilde{Z}_{0}\text{·}A' & w_{2}+\left(a_{4}\tilde{Z}_{14}-\vartheta_{1}^{2}\tilde{Z}_{0}\right)
\end{array}\right|=0.
\]
We view near $\left\{ w_{2}=\tilde{Z}_{0}=Z_{23}=0\right\} $ as a
family of affine quadric surfaces in the coordinates $\left(w_{2},\tilde{Z}_{0},Z_{23}\right)$
parametrized by $a_{4}$ becoming the quadric cone 
\[
\left|\begin{array}{cc}
w_{2}+\vartheta_{1}^{2}\tilde{Z}_{0} & \tilde{Z}_{14}^{2}Z_{23}\\
\tilde{Z}_{0}\text{·}A' & w_{2}-\vartheta_{1}^{2}\tilde{Z}_{0}
\end{array}\right|=0
\]
over $a_{4}\left(p\right)=0$. The fibers over $\left\{ \tilde{Z}_{0}=0\right\} $
are pairs of disjoint affine lines lying in the same plane for all
$a_{4}\text{\ensuremath{\neq}}0$ and so come from opposite rulings
of the $a_{4}$-fiber and coincide as a double line lying in the $\left\{ w_{2}=0\right\} $-plane
when $a_{4}=0$. On the other hand, the fibers over $\left\{ Z_{23}=0\right\} $
are pairs of distinct intersecting affine lines for all $a_{4}$ where
the point of intersection converges to $\left\{ w_{2}=\tilde{Z}_{0}=0\right\} $
as $a_{4}$ goes to zero. Thus the exceptional curve of either small
resolutions will separate the two proper liftings of $\left\{ Z_{23}=0\right\} $
entirely when $a_{4}\left(p\right)=0$. Furthermore the scheme $\left\{ w_{2}=\tilde{Z}_{0}^{2}=0\right\} $
has intersection number one with the exceptional curve, the point
being neither of the points of intersection of the the exceptional
curve with the the two proper liftings of $\left\{ Z_{23}=0\right\} $
.

\subsection{Root configuration over points $p\in S_{\mathrm{GUT}}$}

\subsubsection{Over a general point $p\in\tilde{S}_{\mathrm{GUT}}$ }

Since \eqref{eq:delta2} has no solution on $\left\{ \tilde{Z}_{14}=0\right\} $
therefore $\left\{ \tilde{Z}_{14}\left(p\right)=0\right\} $ cannot
intersect the branch locus $\tilde{\Delta}$. So one concludes that
\[
\tilde{W}_{4}\times_{B_{3}}\left\{ \tilde{Z}_{14}\left(p\right)=0\right\} 
\]
splits into two disjoint components that we denote as $D_{1}\left(p\right)$
(meeting the section $\left(\zeta\right)$) and $D_{4}\left(p\right)$
(meeting the section $\left(\tau\right)$).

$\tilde{\Delta}\cap\left\{ Z_{23}\left(p\right)=0\right\} $ is given
by the equation 
\[
0=\left(\vartheta_{1}\vartheta_{2}+a_{4}\tilde{Z}_{14}\right)^{2}.
\]
Thus 
\[
\tilde{W}_{4}\times_{B_{3}}\left\{ Z_{23}=0\right\} 
\]
consists of two components that cross over the surface $\left\{ \vartheta_{1}\vartheta_{2}+a_{4}\tilde{Z}_{14}=Z_{23}=0\right\} \subseteq\tilde{Q}$.
We have designated these two components as $D_{2}\left(p\right)$
and $D_{3}\left(p\right)$. Furthermore 
\[
C_{\mathbf{\bar{5}}}^{\left(44\right)}\subseteq D_{2}\cap D_{3}.
\]

By \eqref{eq:delta2} $\tilde{\Delta}\cap\left\{ Z_{0}=0\right\} $
is given on $\left\{ Z_{0}=0\right\} $ by the equation 
\[
0=\left(\vartheta_{1}\vartheta_{2}+a_{4}\tilde{Z}_{14}\right)^{2}+4a_{5}a_{420}\tilde{Z}_{14}^{2}.
\]
Thus 
\[
D_{0}\left(p\right):=\tilde{W}_{4}\times_{B_{3}}\left\{ Z_{0}\left(p\right)=0\right\} 
\]
is an irreducible rational curve meeting $D_{1}\left(p\right)$ and
$D_{4}\left(p\right)$, each in a single point. By \eqref{eq:6},
$\left\{ Z_{0}\left(p\right)=0\right\} $ does not meet $\left\{ Z_{23}\left(p\right)=0\right\} $
unless $a_{5}\left(p\right)=0$, so over a general $p\in\tilde{S}_{\mathrm{GUT}}$
, the components of the fiber of $\tilde{W}_{4}/B_{3}$ configure
themselves as the extended Dynkin diagram of $SU\left(5\right)$.

\subsubsection{Over a general point $p\in\Sigma_{\mathbf{10}}^{\left(4\right)}=S_{\mathrm{GUT}}\cap\left\{ a_{5}=0\right\} $ }

On $Q^{\left(2\right)}\times_{B_{3}}\left\{ a_{5}=0\right\} $. 
\[
\vartheta_{2}=-\vartheta_{1}\tilde{Z}_{0}
\]
and the chain configuration becomes 
\[
\left\{ \tilde{Z}_{14}\left(p\right)=0\right\} \cup\left\{ Z_{23}\left(p\right)=0\right\} \cup\left\{ \tilde{Z}_{0}\left(p\right)=0\right\} .
\]
$\Delta^{\left(2\right)}=\tilde{\Delta}$ does not intersect $\left\{ \tilde{Z}_{14}\left(p\right)=a_{5}\left(p\right)=0\right\} $
and does not intersect $\left\{ \tilde{Z}_{0}\left(p\right)=a_{5}\left(p\right)=0\right\} $
since, in general, $a_{4}\text{\ensuremath{\neq}}0$. $\Delta^{\left(2\right)}=\tilde{\Delta}$
intersects $\left\{ Z_{23}=0\right\} $ when $\left(\vartheta_{1}\vartheta_{2}+a_{4}\tilde{Z}_{14}\right)^{2}$,
or equivalently $\left(-\vartheta_{1}^{2}\tilde{Z}_{0}+a_{4}\tilde{Z}_{14}\right)^{2}$.
Thus $D_{2}\left(p\right)$ meets $D_{3}\left(p\right)$ in such a
way that $D_{1}\left(p\right)$ only attaches to $D_{2}\left(p\right)$
and $\left\{ \tilde{Z}_{0}\left(p\right)=0\right\} \times_{B_{3}}\tilde{W}_{4}$
has two disjoint components, one attaching only to $D_{2}\left(p\right)$
and the other attaching only to $D_{3}\left(p\right)$. So over a
general $p\in S_{\mathrm{GUT}}\cap\left\{ a_{5}=0\right\} $ , the
components of the fiber of $\tilde{W}_{4}/B_{3}$ configure themselves
as the extended Dynkin diagram of $SO\left(10\right)$.

\subsubsection{Over a general point $p\in S_{\mathrm{GUT}}$ in the image $\Sigma_{\mathbf{\bar{5}}}^{\left(44\right)}$
of $C_{\mathbf{\bar{5}}}^{\left(44\right)}$}

On this locus either of the two small resolutions of the singular
locus $C_{\mathbf{\bar{5}}}^{\left(44\right)}\subseteq W_{4}^{\left(2\right)}$
inserts an exceptional $\mathbb{P}^{1}$, one via the specialization
\[
D_{2}\left(p\right)\rightsquigarrow D_{2}'\left(p\right)+\bar{E}_{23}\left(p\right)
\]
and the other via the specialization 
\[
D_{3}\left(p\right)\rightsquigarrow D_{3}'+E_{23}\left(p\right).
\]
Both configure the fiber of $\tilde{W}_{4}/B_{3}$ over $p$ as the
extended Dynkin diagram of $SU\left(6\right)$.

\subsubsection{Over a general point $p\in\Sigma_{\mathbf{\bar{5}}}^{\left(41\right)}=S_{\mathrm{GUT}}\cap\left\{ a_{420}=0\right\} $}

Over 
\[
\left\{ Z_{0}\left(p\right)=0\right\} \cup\left\{ \tilde{Z}_{14}\left(p\right)=0\right\} \cup\left\{ Z_{23}\left(p\right)=0\right\} 
\]
the branch locus \eqref{eq:delta2} does not intersect $\left\{ \tilde{Z}_{14}=0\right\} $,
intersects $\left\{ Z_{23}=0\right\} $ when $\left(\vartheta_{1}\frac{\vartheta_{2}}{\tilde{Z}_{14}}+a_{4}\right)^{2}=0$,
and intersects $\left\{ Z_{0}=0\right\} $ when $\left(a_{5}\frac{\vartheta_{1}}{Z_{14}}+a_{4}\right)^{2}=0$.
So again the fiber of $\tilde{W}_{4}/B_{3}$ over $p$ configures
as the extended Dynkin diagram of $SU\left(6\right)$. 
\begin{lem}
Over each point $p\in\left\{ z=0,\,a_{5}\text{\ensuremath{\neq}}0\right\} $
that does not lie in $\Sigma_{\mathbf{\bar{5}}}^{\left(44\right)}\cup\Sigma_{\mathbf{\bar{5}}}^{\left(41\right)}$,
the fibers 
\[
D_{0}\left(p\right)\cup D_{1}\left(p\right)\cup D_{2}\left(p\right)\cup D_{3}\left(p\right)\cup D_{4}\left(p\right)
\]
of $\tilde{W}_{4}\times_{B_{3}}S_{\mathrm{GUT}}$ arrange themselves
in a cyclic intersection configuration corresponding to the extended
Dykin diagram of $SU\left(5\right)$. If $p\in\Sigma_{\mathbf{\bar{5}}}^{\left(44\right)}$,
$D_{2}\left(p\right)$ and $D_{3}\left(p\right)$ become separated
by a new component lying in the crepant resolution of the locus $C_{\mathbf{\bar{5}}}^{\left(44\right)}$
that we declare as the 'new root' yielding a configuration corresponding
to the extended Dykin diagram of $SU\left(6\right)$. This new root
either emerges as $\bar{E}_{23}\left(p\right)$ from the specialization
\[
D_{2}'\left(p\right)+\bar{E}_{23}\left(p\right)
\]
of $D_{2}\left(p\right)$ or as $E_{23}\left(p\right)$ from the specialization
\[
E\left(p\right)+D_{3}'\left(p\right)
\]
of $D_{3}\left(p\right)$ depending on the choice of small resolution
of $C_{\mathbf{\bar{5}}}^{\left(44\right)}$.

If $p\in\Sigma_{\mathbf{\bar{5}}}^{\left(41\right)}$, $D_{0}\left(p\right)$
is split in two by a contracting circle 'half-way' between the intersection
points $D_{0}\cap\left(D_{1}\cup D_{4}\right)$ again $C_{\mathbf{\bar{5}}}^{\left(44\right)}$
yielding a cyclic configuration 
\[
D_{40}\left(p\right)\cup D_{01}\left(p\right)\cup D_{1}\left(p\right)\cup D_{2}\left(p\right)\cup D_{3}\left(p\right)\cup D_{4}\left(p\right)
\]
corresponding to the extended Dykin diagram of $SU\left(6\right)$.

If $p\in\Sigma_{\mathbf{\bar{5}}}^{\left(41\right)}\cap\Sigma_{\mathbf{\bar{5}}}^{\left(44\right)}$,
both of the two above splittings occur simultaneously yielding a configuration
corresponding to the extended Dykin diagram of $SU\left(7\right)$. 
\end{lem}

The situation over 
\[
\Sigma_{\mathbf{10}}^{\left(4\right)}:=\left\{ z=a_{5}=0\right\} \subseteq B_{3}
\]
is more complex. $\left\{ Z_{0}\left(p\right)=0\right\} $ contains
the center of the second modification, that is,

\begin{equation}
Z_{0}\left(p\right)=\tilde{Z}_{0}\left(p\right)Z_{23}\left(p\right).\label{eq:8}
\end{equation}
Furthermore the intersection of the branch locus $\tilde{\Delta}=\Delta^{\left(2\right)}$
with $\left\{ Z_{0}\left(p\right)=0\right\} $ deposits entirely on
the component $\left\{ Z_{23}\left(p\right)=0\right\} $ as a point
of multiplicity two not on either of the other components unless additionally
$a_{4}=0$ where it becomes the entire fiber $\left\{ \tilde{Z}_{0}^{2}\left(p\right)=0\right\} $
. 
\begin{lem}
\label{lem:i)-Over-points}i) Over points of $\left\{ z=a_{5}=0\right\} \cap\left\{ a_{4}a_{420}\text{\ensuremath{\neq}}0\right\} \subseteq B_{3}$
the fiber of $\tilde{Q}/B_{3}$ consists of a chain of three $\mathbb{P}^{1}$'s,
with the property that the branch locus $\tilde{\Delta}$ meets only
the middle component, namely $\left\{ Z_{23}=0\right\} $ which it
meets in one point. Thus the fiber of $\tilde{W}_{4}/B_{3}$ is a
tree of $\mathbb{P}^{1}$'s whose intersection configuration is that
of the extended Dynkin diagram of $SO\left(10\right)$. The 'new root'
is obtained by $D_{0}\left(p\right)$ becoming reducible as 
\begin{equation}
D_{0}\left(p\right)=D_{02}\left(p\right)+D_{03}\left(p\right)\label{eq:fact2}
\end{equation}
with $D_{03}\left(p\right)$ meeting only $D_{2}\left(p\right)$ and
$D_{03}\left(p\right)$ meeting only $D_{3}\left(p\right)$.

ii) If $z=a_{5}=a_{4}=0$ then the curve $C_{\mathbf{\bar{5}}}^{\left(44\right)}\subseteq\Delta^{\left(2\right)}$
passes over $\left\{ z=a_{5}=a_{4}=0\right\} \subseteq B_{3}$ contributing
a last component to the fiber of $\tilde{W}_{4}/B_{3}$. This last
component meets the component $\left\{ Z_{23}=0\right\} $ at the
point where $\tilde{Z}_{0}=0$. Thus the fiber of $\tilde{W}_{4}/B_{3}$
contains a tree of $\mathbb{P}^{1}$'s configured as the Dynkin diagram
of $E_{6}$. In fact the scheme-theoretic fiber, where the 'short
arm' occurs with multiplicity two is a flat specialization of the
tree given by the extended Dynkin diagram of $E_{6}$.

iii) If $z=a_{5}=a_{420}=0$ , then by \eqref{eq:secsec} $a_{3}$
also vanishes. So the curve $C_{\mathbf{\bar{5}}}^{\left(44\right)}\subseteq\Delta^{\left(2\right)}$
in Subsection 7.2.3 of \cite{Clemens-3} passes over $\left\{ z=a_{5}=a_{3}=0\right\} \subseteq B_{3}$
contributing a last component to the fiber of $\tilde{W}_{4}/B_{3}$.
This last component meets the component $\left\{ Z_{23}=0\right\} $
at the point where $\vartheta_{1}\tilde{\vartheta}_{2}\tilde{Z}_{0}+a_{4}\tilde{Z}_{14}=0$.
Thus the fiber of $\tilde{W}_{4}/B_{3}$ is a tree of $\mathbb{P}^{1}$'s
whose intersection configuration is that of the extended Dynkin diagram
of $SO\left(12\right)$. 
\end{lem}

\section{$U\left(1\right)_{X}$-charges}

In our model in \cite{Clemens-3} the $U\left(1\right)_{X}$-section
is derived from the line bundle 
\[
\mathcal{O}_{W_{4}}\left(\left(\zeta\right)-\left(\tau\right)\right)
\]
where $\left(\tau\right)$ is the divisor associated with the section
$\tau$ and $\left(\zeta\right)$ is the divisor associated with the
section $\zeta$. Lifting $\left(\zeta\right)-\left(\tau\right)$
to the divisor $\left(\tilde{\zeta}\right)-\left(\tilde{\tau}\right)$
in the desingularization $\tilde{W}_{4}$ we obtain its $U\left(1\right)_{X}$-action
by normalizing it so as to have zero intersection with all components
of a general fiber over $S_{\mathrm{GUT}}$. Since, for $p\in S_{\mathrm{GUT}}$,
the section $\left(\zeta\right)$ meets $D_{1}\left(p\right)$ and
the section $\left(\tau\right)$ meets $D_{4}\left(p\right)$, we
obtain the divisor 
\begin{equation}
\left[U\left(1\right)_{X}\right]:\,\,5\left(\left(\tilde{\zeta}\right)-\left(\tilde{\tau}\right)\right)+\left(3D_{1}+D_{2}-D_{3}-3D_{4}\right).\label{eq:U}
\end{equation}
This divisor is orthogonal to all four of the $D_{j}\left(p\right)$
and therefore supports the $U\left(1\right)_{X}$-action.

\subsection{$U\left(1\right)_{X}$-charge on $\Sigma_{\mathbf{10}}^{\left(4\right)}\subseteq D_{4}$}

Over $\Sigma_{\mathbf{10}}^{\left(4\right)}\subseteq B_{3}$ the new
root is given by the fact that $\left\{ Z_{0}\left(p\right)=0\right\} $
becomes reducible via the factorization $Z_{0}=Z_{23}\tilde{Z}_{0}$.
Furthermore the two branch-points of the irreducible branched double
cover $D_{0}\left(p\right)=\left\{ p\right\} \times_{B_{3}}\tilde{W}_{4}-\sum_{j=1}^{4}D_{j}\left(p\right)$
over $\left\{ Z_{0}\left(p\right)=0\right\} $ specializes to a single
double-point lying on $\left\{ Z_{23}=0\right\} $. As we saw in Lemma
\ref{lem:i)-Over-points} above this splits $D_{0}\left(p\right)$
into two components $D_{02}\left(p\right)+D_{03}$. Either of these
can be chosen as the 'new root.' $D_{02}\left(p\right)$ has one intersection
with $D_{2}$ and zero intersection with all other roots and with
$\left(\zeta\right)$ and $\left(\tau\right)$ since the two sections
intersect $D_{1}\left(p\right)$ and $D_{4}\left(p\right)$ respectively.
$D_{03}\left(p\right)$ has one intersection with $D_{3}$ and zero
intersection with all other roots and with $\left(\zeta\right)$ and
$\left(\tau\right)$.

On the other hand, over the matter curve $\Sigma_{\mathbf{\bar{5}}}^{\left(41\right)}$,
$D_{0}\left(p\right)$ again becomes reducible but this time $\left\{ Z_{0}\left(p\right)=0\right\} $
does not, rather the branch-points of $D_{0}\left(p\right)$ over
$\left\{ Z_{0}\left(p\right)=0\right\} $ come together to give 
\[
D_{0}\left(p\right)=D_{01}\left(p\right)+D_{40}\left(p\right)
\]
where $D_{01}\left(p\right)$ has one intersection with $D_{1}$ and
zero intersection with all other roots and $D_{40}\left(p\right)$
has one intersection with $D_{4}$ and zero intersection with all
other roots. $D_{01}\left(p\right)$ also misses $\left(\zeta\right)$
and $\left(\tau\right)$ even though $D_{01}\left(p\right)$ and $\left(\zeta\right)$
both meet $D_{1}\left(p\right)$. We designate $D_{01}\left(p\right)$
as the new root and so the matter curve $\Sigma_{\mathbf{\bar{5}}}^{\left(41\right)}$
has $U\left(1\right)_{X}$-charge $+3$.

Finally over the Higgs curve $\Sigma_{\mathbf{\bar{5}}}^{\left(44\right)}$
the new root $E_{23}\left(p\right)$ arises from a splitting of $D_{2}\left(p\right)$
or $D_{3}\left(p\right)$ (depending on which of the two small resolutions
of the nodal locus of $W_{4}^{\left(2\right)}$ we choose). If $E_{23}$
is incorporated into $D_{3}=D'_{3}+E_{23}$ we have 
\[
E_{23}\text{·}\left[U\left(1\right)_{X}\right]=E_{23}\text{·}\left(D_{2}-E_{23}-D'_{3}\right)=+1-\left(-2\right)-1=+2
\]
and $D'_{3}\text{·}\left[U\left(1\right)_{X}\right]=-2$, whereas
if $E_{23}$ is incorporated into $D_{2}=D'_{2}+\bar{E}_{23}$ we
have 
\[
\bar{E}_{23}\text{·}\left[U\left(1\right)_{X}\right]=\bar{E}_{23}\text{·}\left(D'_{2}+\bar{E}_{23}-D{}_{3}\right)=-1+\left(-2\right)+1=-2
\]
and $D'_{2}\text{·}\left[U\left(1\right)_{X}\right]=+2$. Here we
can choose either small resolution since in either we have a curve
with charge $+2$ and a curve $-2$, either of which can be designated
as the 'new root' as needed. We summarize charges as follows:\smallskip{}
 
\begin{center}
\begin{tabular}{|c|c|c|}
\hline 
New root/component  & Curve/$SU\left(5\right)_{gauge}$-representation  & $\left[U\left(1\right)_{X}\right]$-charge\tabularnewline
\hline 
\hline 
$D_{02}$ meets $D_{2}$  & $\Sigma_{\mathbf{10}}^{\left(4\right)}$  & $+1$\tabularnewline
\hline 
$D_{03}$ meets $D_{3}$  & $\Sigma_{\mathbf{10}}^{\left(4\right)}$  & $-1$\tabularnewline
\hline 
$D_{01}$ meets $D_{1}$  & $\Sigma_{\mathbf{\bar{5}}}^{\left(41\right)}$  & $+3$\tabularnewline
\hline 
$D_{40}$ meets $D_{4}$  & $\Sigma_{\mathbf{\bar{5}}}^{\left(41\right)}$  & $-3$\tabularnewline
\hline 
$D'_{3},\,\bar{E}_{23}$  & $\Sigma_{\mathbf{\bar{5}}}^{\left(44\right)}$  & $-2$\tabularnewline
\hline 
$D'_{2},\,E_{23}$  & $\Sigma_{\mathbf{5}}^{\left(44\right)}$  & $+2$\tabularnewline
\hline 
\end{tabular}\smallskip{}
 
\par\end{center}

Thus there are two possible choices for the assignment of charges
to the three curves that are compatible with the restriction of $E_{8}\overset{Ad_{E_{8}}}{\longrightarrow}Aut\left(e_{8}\right)$
to $\left(SU\left(5\right)\right)_{gauge}\times\left(SU\left(4\right)_{Higgs}\times U\left(1\right)_{X}\right)$
and with the Yukawa couplings 
\[
\left(\mathbf{10}_{M},\mathbf{10}_{M},\mathbf{5}_{H}\right)\,and\,\left(\mathbf{10}_{M},\bar{\mathbf{5}}_{M},\bar{\mathbf{5}}_{H}\right)
\]

One decomposes as in (88) of \cite{Blumenhagen} into 
\begin{equation}
\begin{array}{c}
\left(\mathbf{1},\mathbf{15}\right)_{0}\oplus\\
\left(\mathbf{1},\mathbf{1}\right)_{0}\oplus\left(\mathbf{10},\mathbf{1}\right)_{-4}\oplus\left(\mathbf{\overline{10}},\mathbf{1}\right)_{4}\oplus\left(\mathbf{24},\mathbf{1}\right)_{0}\\
\oplus\left(\mathbf{\mathbf{1}},\mathbf{4}\right)_{5}\oplus\left(\mathbf{\overline{5}},\mathbf{4}\right)_{-3}\oplus\left(\mathbf{10},\mathbf{4}\right)_{1}\\
\oplus\left(\mathbf{1},\mathbf{\overline{4}}\right)_{-5}\oplus\left(\mathbf{5},\mathbf{\bar{4}}\right)_{3}\oplus\left(\overline{\mathbf{10}},\mathbf{4}\right)_{-1}\\
\oplus\left(\mathbf{\mathbf{5}},\mathbf{6}\right)_{-2}\oplus\left(\mathbf{\overline{5}},\mathbf{6}\right)_{2}
\end{array}\label{eq:Ad}
\end{equation}
where $u\in U\left(1\right)_{X}$ acts on the representation space
$\left(\right)_{c}$ as the character representation 
\[
u\left(v\right)\mapsto u^{c}\text{·}v
\]
and the other decomposes into 
\begin{equation}
\begin{array}{c}
\left(\mathbf{1},\mathbf{15}\right)_{0}\oplus\\
\left(\mathbf{1},\mathbf{1}\right)_{0}\oplus\left(\mathbf{10},\mathbf{1}\right)_{4}\oplus\left(\mathbf{\overline{10}},\mathbf{1}\right)_{-4}\oplus\left(\mathbf{24},\mathbf{1}\right)_{0}\\
\oplus\left(\mathbf{\mathbf{1}},\mathbf{4}\right)_{-5}\oplus\left(\mathbf{\overline{5}},\mathbf{4}\right)_{3}\oplus\left(\mathbf{10},\mathbf{4}\right)_{-1}\\
\oplus\left(\mathbf{1},\mathbf{\overline{4}}\right)_{5}\oplus\left(\mathbf{5},\mathbf{\bar{4}}\right)_{-3}\oplus\left(\overline{\mathbf{10}},\mathbf{4}\right)_{1}\\
\oplus\left(\mathbf{\mathbf{5}},\mathbf{6}\right)_{2}\oplus\left(\mathbf{\overline{5}},\mathbf{6}\right)_{-2}.
\end{array}\label{eq:Ad'}
\end{equation}

\subsection{Right handed neutrinos}

Now the matter curve $\Sigma_{\mathbf{\bar{5}}}^{\left(41\right)}$
is given by 
\[
a_{420}=\left(t+z\right)=0
\]
lying inside the surface of intersection of $\mathcal{D}^{\left(4\right)}$
over $\left\{ a_{420}=0\right\} \subseteq B_{3}$ . In fact the restriction
of $\mathcal{D}^{\left(4\right)}$ factors over $\left\{ a_{420}=0\right\} $
and we have the equation 
\begin{equation}
\begin{array}{c}
\frac{a_{5}t^{4}+a_{54}t^{3}z-a_{20}t^{2}z^{2}-a_{0}t^{3}\left(t+z\right)}{t+\tilde{z}}=\\
\frac{a_{5}t^{4}+a_{54}t^{3}z+a_{4}t^{2}z^{2}-a_{0}t^{3}\left(t+z\right)}{t+z}=\\
a_{5}t^{3}+a_{4}t^{2}z-a_{0}z^{3}=0
\end{array}\label{eq:1}
\end{equation}
for this residual surface. From §7.4 in \cite{Clemens-3} we have
the equation 
\[
t=\frac{y}{x}=\frac{\left(\vartheta_{0}-a_{420}\right)-z}{2}
\]
so that over $\left\{ a_{420}=0\right\} $ we have the correspondence
\[
\begin{array}{c}
a_{5}t^{3}+a_{4}t^{2}z-a_{0}z^{3}=t^{2}\left(a_{5}t+a_{4}z\right)-a_{0}z^{3}=\\
\left(\frac{\vartheta_{0}-z}{2}\right)^{2}\left(a_{5}\left(\frac{\vartheta_{0}-z}{2}\right)+a_{4}z\right)-a_{0}z^{3}.
\end{array}
\]
The proper transform of the surface in $Q^{\left(1\right)}\times_{B_{3}}\left\{ a_{420}=0\right\} $
defined by this last polynomial has equation 
\begin{equation}
Z_{0}=a_{5}\left(\frac{\vartheta_{1}-Z_{14}}{2}\right)^{3}+a_{4}Z_{14}\left(\frac{\vartheta_{1}-Z_{14}}{2}\right)^{2}-a_{0}Z_{14}^{3}=0\label{eq:13}
\end{equation}
that does not meet $\left\{ Z_{14}=0\right\} $ unless $a_{5}$ also
vanishes since $\vartheta_{1}$ and $Z_{14}$ have no common zeros.
Now for $p\in\left\{ z=a_{420}=0\right\} $, $D_{0}\left(p\right)=D_{01}\left(p\right)+D_{40}\left(p\right)$
neither component of which connect with $D_{2}\left(p\right)+D_{3}\left(p\right)$
unless $a_{5}\left(p\right)$ also vanishes. By \eqref{eq:DSp-1},
$D_{01}\left(p\right)$ crosses $D_{40}\left(p\right)$ when $\frac{\vartheta_{1}}{Z_{14}}=-\frac{a_{4}}{a_{5}}$
whereas the locus \eqref{eq:13} is given by the locus 
\[
Z_{0}=\frac{a_{5}}{8}\left(\frac{\vartheta_{1}}{Z_{14}}-1\right)^{3}+\frac{a_{4}}{4}\left(\frac{\vartheta_{1}}{Z_{14}}-1\right)^{2}-a_{0}=0
\]
Since the cubic \eqref{eq:13} lies in the spectral divisor, it lifts
into $D_{40}$ and meets no other component of the fiber as long as
$a_{5}\text{\ensuremath{\neq}}0$. We denote the closure of this locus
in $D_{40}$ as 
\[
\Gamma_{0}\subseteq\left\{ z=a_{420}=0\right\} \times_{B_{3}}D_{04}\subseteq W_{4}^{\left(1\right)}.
\]

If additionally $a_{5}\left(p\right)=0$ then, by allowably general
choices of the forms $a_{j}$, $a_{4}\left(p\right)\text{\ensuremath{\neq}}0$
and solutions to \eqref{eq:13} specialize to 
\[
\begin{array}{c}
Z_{0}\left(p\right)=Z_{14}\left(p\right)=0\\
Z_{0}\left(p\right)=0,\,\left(\frac{\vartheta_{1}}{Z_{14}}\left(p\right)-1\right)^{2}=\frac{4a_{0}\left(p\right)}{a_{4}\left(p\right)}.
\end{array}
\]
Thus $\Gamma_{0}$ meets $\left\{ Z_{14}=0\right\} $ simply over
$a_{5}\left(p\right)=0$.

Now the transform of \eqref{eq:13} in $Q^{\left(2\right)}\times_{B_{3}}\left\{ Z_{0}=a_{420}=0\right\} $
becomes 
\[
\left(\frac{-\tilde{Z}_{14}Z_{23}}{2}\right)^{2}\left(a_{5}\left(\frac{-\tilde{Z}_{14}Z_{23}}{2}\right)+a_{4}\tilde{Z}_{14}Z_{23}\right)-a_{0}\tilde{Z}_{14}^{3}Z_{23}^{3}=0
\]
or alternatively 
\[
\left(\frac{a_{5}-\left(\tilde{Z}_{14}+\vartheta_{2}\right)Z_{23}}{2}\right)^{2}\left(a_{5}\left(\frac{a_{5}-\left(\tilde{Z}_{14}+\vartheta_{2}\right)Z_{23}}{2}\right)+a_{4}\tilde{Z}_{14}Z_{23}\right)-a_{0}\tilde{Z}_{14}^{3}Z_{23}^{3}=0.
\]

Also on $\left\{ Z_{0}=0\right\} $ we can use the second form of
the equation to obtain a curve 
\[
\left\{ Z_{0}=a_{420}=\left(\frac{a_{5}-\left(\tilde{Z}_{14}+\vartheta_{2}\right)Z_{23}}{2}\right)^{2}\left(a_{5}\left(\frac{a_{5}-\left(\tilde{Z}_{14}+\vartheta_{2}\right)Z_{23}}{2}\right)+a_{4}\tilde{Z}_{14}Z_{23}\right)-a_{0}\tilde{Z}_{14}^{3}Z_{23}^{3}=0\right\} 
\]
whose lifting into $D_{04}\subseteq W_{4}^{\left(2\right)}$ we designate
as $\tilde{\Gamma}_{0}$. Now $\tilde{\Gamma}_{0}$ is disjoint from
the support of $\left[U\left(1\right)_{X}\right]$ except over the
point $\left\{ z\left(p\right)=a_{5}\left(p\right)=a_{420}\left(p\right)=0\right\} $
where it meets the remaining components over $\left\{ \tilde{Z}_{14}Z_{23}=0\right\} $.
Depending on the choice of small resolution over the Higgs curve,
define the right handed neutrino curve as 
\[
\tilde{\Gamma}\equiv\tilde{\Gamma}_{0}+D_{4}\left(p\right)+D_{3}\left(p\right)+\bar{E}_{23}\left(p\right)
\]
or 
\[
\tilde{\Gamma}\equiv\tilde{\Gamma}_{0}+D_{4}\left(p\right)+D'_{3}\left(p\right)
\]
where $p\in\left\{ z=a_{420}=a_{5}=0\right\} \subseteq S_{\mathrm{GUT}}$
and $\tilde{\Gamma}_{0}$ meets $D_{4}$ simply above the point $\left\{ a_{5}=a_{420}=\tilde{Z}_{0}=\tilde{Z}_{14}=0\right\} \subseteq Q^{\left(2\right)}$.

We have the following table of intersection numbers with \eqref{eq:U}:\smallskip{}
 
\begin{center}
\begin{tabular}{|c|c|}
\hline 
$\tilde{\Gamma}$-component  & Intersection number with $\left[U\left(1\right)_{X}\right]$\tabularnewline
\hline 
\hline 
$\tilde{\Gamma}_{0}$  & $-3$\tabularnewline
\hline 
$D_{4}\left(p\right)$  & $-5+\left(-2\right)\left(-3\right)+\left(-1\right)=0$\tabularnewline
\hline 
$D_{3}\left(p\right)$  & $-3+2+1=0$\tabularnewline
\hline 
$\bar{E}_{23}\left(p\right),\,D_{3}'\left(p\right)$  & $-2$\tabularnewline
\hline 
\end{tabular}\smallskip{}
 
\par\end{center}

\begin{flushleft}
Recall now that the choice in the last entry in the table depends
on the choice of small resolution of $C_{\mathbf{\bar{5}}}^{\left(44\right)}$
over $p\in\left\{ z=a_{5}=a_{420}=0\right\} $. For either of the
choices 
\[
\tilde{\Gamma}\text{·}\left[U\left(1\right)_{X}\right]=-5
\]
and, following §4 of \cite{Blumenhagen} the curve $\tilde{\Gamma}$
is the candidate for the the support of right-handed neutrinos. However
the geometric construction forces a Yukawa coupling 
\[
\left(\mathbf{1}_{\tilde{\Gamma}},\bar{\mathbf{5}}_{M},\mathbf{5}_{H}\right)
\]
since the curves $\Sigma_{\mathbf{\bar{5}}}^{\left(41\right)}$ and
$\Sigma_{\mathbf{\bar{5}}}^{\left(44\right)}$ lie in $D_{4}$. Thus
we are forced to assign charge $+3$ to $\Sigma_{\mathbf{\bar{5}}}^{\left(41\right)}$
as in \eqref{eq:Ad'}. 
\par\end{flushleft}

The fact that $\tilde{\Gamma}$ is reducible will force a coupling
\[
\left(\mathbf{\mathbf{1}}_{\tilde{\Gamma}},\mathbf{\mathbf{1}}_{\tilde{\Gamma}},\mathbf{1}_{\tilde{\Lambda}}\right)
\]
involving a curve with $U\left(1\right)_{X}$-charge $+10$. To construct
this curve, consider the curve $\tilde{\Lambda}_{0}=D_{01}\left(p\right)$
for $p\in S_{\mathrm{GUT}}$ such that $a_{420}\left(p\right)=a_{5}\left(p\right)=0$.
Then
\[
\left[U\left(1\right)_{X}\right]\text{·}\tilde{\Lambda}_{0}=\left(-5\right)\left(\tilde{\zeta}\right)\text{·}\tilde{\Lambda}_{0}+3D_{1}\text{·}\tilde{\Lambda}_{0}=8
\]
and so 
\[
\left[U\left(1\right)_{X}\right]\text{·}\left(\tilde{\Lambda}_{0}+D_{1}\left(p\right)+D'_{2}\left(p\right)\right)=10
\]
and
\[
\left[U\left(1\right)_{X}\right]\text{·}\left(\tilde{\Lambda}_{0}+D_{1}\left(p\right)+D{}_{2}\left(p\right)+E_{23}\left(p\right)\right)=10.
\]
Now the choice between 
\[
\tilde{\Lambda}=\tilde{\Lambda}_{0}+D_{1}\left(p\right)+D'_{2}\left(p\right)
\]
or
\[
\tilde{\Lambda}=\tilde{\Lambda}_{0}+D_{1}\left(p\right)+D{}_{2}\left(p\right)+E_{23}\left(p\right)
\]
 must be coordinated with the choice between
\[
\tilde{\Gamma}=\tilde{\Gamma}_{0}+D_{4}\left(p\right)+D_{3}\left(p\right)+\bar{E}_{23}\left(p\right)
\]
or 
\[
\tilde{\Gamma}=\tilde{\Gamma}_{0}+D_{4}\left(p\right)+D'_{3}\left(p\right)
\]
so that $\tilde{\Lambda}$ and $\tilde{\Gamma}$ pass through a common
point.

Charges associated to the $U\left(1\right)_{X}$-action are then given
by the following table:\smallskip{}
 
\begin{center}
\begin{tabular}{|c|c|c|}
\hline 
New root/component  & Curve/$SU\left(5\right)_{gauge}$-representation  & $\left[U\left(1\right)_{X}\right]$-charge\tabularnewline
\hline 
\hline 
$D_{03}$ meets $D_{3}$  & $\Sigma_{\mathbf{10}}^{\left(4\right)}$  & $-1$\tabularnewline
\hline 
$D_{01}$ meets $D_{1}$  & $\Sigma_{\mathbf{\bar{5}}}^{\left(41\right)}$  & $+3$\tabularnewline
\hline 
$D'_{3},\,\bar{E}_{23}$  & $\Sigma_{\mathbf{\bar{5}}}^{\left(44\right)}$  & $-2$\tabularnewline
\hline 
$D'_{2},\,E_{23}$  & $\Sigma_{\mathbf{5}}^{\left(44\right)}$  & $+2$\tabularnewline
\hline 
Right-handed neutrinos  & $\tilde{\Gamma}$  & $-5$\tabularnewline
\hline 
 & $\tilde{\Lambda}$  & $+10$\tabularnewline
\hline 
\end{tabular}
\par\end{center}

\smallskip{}

\section{Yukawa couplings}

\eqref{eq:secsec} then implies that at those points we also have
$a_{3}=0$ since our second section required $-a_{420}=a_{53}$. So
we conclude that over the twelve points $\left\{ a_{5}=a_{420}=z=0\right\} \subseteq B_{3}$
both matter curves, the Higgs curve and the curve that supports the
right-handed neutrinos all intersect.

We next recall from §6.2.1 of \cite{Clemens-2} that the curve $\left\{ a_{5}=z=0\right\} \subseteq B_{3}$
contains two disjoint genus-one curves given by $\left\{ u_{0}v_{0}=0\right\} $
that are interchanged under the $\mathbb{Z}_{2}$-action and each
meets the residual curve in two points on which $a_{4}$ can be chosen
to vanish. It is over these crossing points of components that we
find the 'top' $\left(\mathbf{10}_{M},\mathbf{10}_{M},\mathbf{5}_{H}\right)$-couplings
associated with the $U\left(1\right)_{X}$-action. Furthermore, since
$a_{3}$ also vanishes when $a_{5}=a_{420}=0$ allowing 'bottom' $\left(\mathbf{10}_{M},\bar{\mathbf{5}}_{M},\bar{\mathbf{5}}_{H}\right)$-couplings
associated with the $U\left(1\right)_{X}$ action over $\left\{ a_{5}=a_{420}=0\right\} $.
Next points where the neutrino curve meets the $\bar{\mathbf{5}}_{M}$-curve
and the $\mathbf{5}_{H}$-curve give $\left(\mathbf{1}_{\tilde{\Gamma}},\bar{\mathbf{5}}_{M},\mathbf{5}_{H}\right)$-couplings.
Finally the crossing point of the components $\tilde{\Gamma}_{0}$
and $D_{1}\left(p\right)$ of the neutrino curve $\tilde{\Gamma}$
over $\left\{ z=a_{5}=a_{420}=0\right\} $ allows a $\left(\mathbf{\mathbf{1}}_{\tilde{\Gamma}},\mathbf{\mathbf{1}}_{\tilde{\Gamma}},\mathbf{1}_{\tilde{\Lambda}}\right)$-coupling.
We have the following table: 
\begin{center}
\smallskip{}
 
\par\end{center}

\begin{center}
\begin{tabular}{|c|c|}
\hline 
Coupling  & Charge\tabularnewline
\hline 
\hline 
$\left(\mathbf{10}_{M},\mathbf{10}_{M},\mathbf{5}_{H}\right)$  & $\left(-1\right)+\left(-1\right)+\left(+2\right)=0$\tabularnewline
\hline 
$\left(\mathbf{10}_{M},\bar{\mathbf{5}}_{M},\bar{\mathbf{5}}_{H}\right)$  & $\left(-1\right)+\left(+3\right)+\left(-2\right)=0$\tabularnewline
\hline 
$\left(\mathbf{1}_{\tilde{\Gamma}},\bar{\mathbf{5}}_{M},\mathbf{5}_{H}\right)$  & $\left(-5\right)+\left(+3\right)+\left(+2\right)=0$\tabularnewline
\hline 
$\left(\mathbf{\mathbf{1}}_{\tilde{\Gamma}},\mathbf{\mathbf{1}}_{\tilde{\Gamma}},\mathbf{1}_{\tilde{\Lambda}}\right)$  & $\left(-5\right)+\left(-5\right)+\left(+10\right)=0$\tabularnewline
\hline 
\end{tabular}
\par\end{center}

\section{$\mathbb{Z}_{2}$-quotient}

The $U\left(1\right)_{X}$-charge $c$ acts on the representations
\eqref{eq:Ad} as the character $u\mapsto u^{c}$. Under the $\mathbb{Z}_{2}$-action,
the section $\tau$ is interchanged with the inherited section $\zeta$,
so the line bundle $\mathcal{O}_{W_{4}}\left(\left(\tau\right)-\left(\zeta\right)\right)$
is carried to its inverse. Roots are sent to their negatives by the
reversal of the choice of positive Weyl chamber but that reversal
is undone by the first factor of the composite $\mathbb{Z}_{2}$-action
that sends $y$ to $-y$ but does \textit{not} interchange $\tau$
with $\zeta$. It is the second factor of the $\mathbb{Z}_{2}$-action,
namely the fiberwise translation by $\mathrm{image}\left(\tau\right)-\mathrm{image}\left(\zeta\right)$
that carries the line bundle $\mathcal{O}_{W_{4}}\left(\left(\tau\right)-\left(\zeta\right)\right)$
to its inverse. (See Sections 9 and 10.1 of \cite{Clemens-3}.) Therefore
\[
\left[U\left(1\right)_{X}\right]\Rightarrow-\left[U\left(1\right)_{X}\right].
\]
Since the composite $\mathbb{Z}_{2}$-action on the $U\left(1\right)_{X}$
lying in the maximal torus of the complex algebraic group $E_{8}^{\mathbb{C}}$
is by complex conjugation, the only possible non-zero $U\left(1\right)_{X}$-charges
$c$ that are $\mathbb{Z}_{2}$\textendash invariant are those such
that 
\[
u^{c}=u^{-c}.
\]

Therefore requiring that 
\begin{equation}
u^{c}=u^{-c}\label{eq:break}
\end{equation}
implies 
\[
u^{2c}=1,
\]
that is 
\[
u^{c}=e^{m\pi i}=\text{\textpm}1.
\]
Said otherwise, the requirement \eqref{eq:break} breaks $U\left(1\right)_{X}$-symmery
to $\mathbb{Z}_{2}$-symmetry. That is, as mentioned in \eqref{eq:break}
above, descent to the $\mathbb{Z}_{2}$-quotient breaks $U\left(1\right)_{X}$-symmetry
to $\mathbb{Z}_{2}$-symmetry. Thus the right-handed neutrinos can
obtain a Majorana mass at the compactification scale.

Furthermore, the $\mathbb{Z}_{2}$-action moves the spectral divisor
\[
\tilde{\mathcal{C}}_{Higgs}^{\left(4\right)}\subseteq\tilde{W}_{4}
\]
to the opposite component of 
\[
\tilde{\mathcal{C}}_{Higgs}^{\left(4\right)}\times_{\tilde{Q}}\tilde{W}_{4},
\]
namely to the component whose intersection with $\tilde{W}_{4}\times_{B_{3}}S_{\mathrm{GUT}}$
lies in $D_{1}$. But this 'opposite' component is, in fact, the spectral
divisor with respect to the opposite choice of positive Weyl chamber
and its associated opposite (or 'flopped') Brieskorn-Grothendieck
equivariant crepant resolution as explained in \cite{Clemens-1}.
Since $\tilde{\Gamma}$ and $\tilde{\Lambda}$ are defined entirely
with respect to their relationship with the spectral variety for their
respective Brieskorn-Grothendieck equivariant crepant resolutions,
both $\tilde{\Gamma}$ and $\tilde{\Lambda}$ will be taken to themselves
under the $\mathbb{Z}_{2}$-action.

In summary, the $\left(4+1\right)$-spectral equation breaks $E_{8}$-symmetry
to 
\[
SU\left(5\right)_{gauge}\times U\left(1\right)_{X}\subseteq SO\left(10\right)
\]
and the $\mathbb{Z}_{2}$-action breaks $U\left(1\right)_{X}$, leaving
only $SU\left(5\right)_{gauge}\times\mathbb{Z}_{2}$-symmetry before
wrapping the Wilson line. In particular, the right-handed neutrinos
discussed above can in principle obtain a large Majorana mass near
the GUT/compactification scale. Thus the theory of \cite{Clemens-3}
contains three families of quarks and leptons and one pair of Higgs
doublets (after $\mathrm{GUT}$ symmetry breaking via the Wilson line).
Note, the theory also includes a complete MSSM twin sector. The $\mathbb{Z}_{2}$-symmetry
is identified as matter parity. However, in addition as explained
in \cite{Clemens-2}, the theory has an asymptotic $\mathbb{Z}_{4}^{\mathbf{R}}$-symmetry
which forbids dimension $4$ and $5$ baryon-violating operators as
well as the Higgs $\mu$ term. Finally the theory has non-trivial
Yukawa couplings located at co-dimension $3$ singularities.

Further analysis is necessary to determine the $3\times3$ Yukawa
coupling matrices. We also want to analyze the MSSM twin world in
order to address the question of relative scales between the visible
and twin sectors and to determine whether or not there are any possible
portals to the twin sector. Finally the issue of supersymmetry breaking
must be addressed.

\end{document}